%Paper: hep-ph/9502279
%From: RCNP Shoichi Sasaki <ssasaki@miho.rcnp.osaka-u.ac.jp>
%Date: Sat, 11 Feb 1995 16:00:00 +0900

\input PHYZZX

%%%%%%%%%%%%%%%%%%% title page %%%%%%%%%%%%%%%%%%%%%%%%%%%%

\titlepage

\title{
Dual Higgs Mechanism for Quarks in Hadrons
\foot{
To appear in Proc. of the YITP Workshop on
``From Hadronic Matter to Quark Matter: Evolving
View of Hadronic Matter'', YITP Kyoto Japan,
November 1994,
Prog.~Theor.~Phys.~(Supplement).
}
}

\author{
H.~Suganuma$^{\rm a}$, S.~Sasaki$^{\rm b}$, H.~Toki$^{\rm b}$
and H.~Ichie$^{\rm c}$}

\address{ {\rm a)}
The Institute of Physical and Chemical Research (RIKEN)
\break
Hirosawa 2-1, Wako, Saitama 351-01, Japan
}

\address{ {\rm b)}
Research Center for Nuclear Physics (RCNP), Osaka University
\break
Mihogaoka 10-1, Ibaraki, Osaka 567, Japan
}

\address{{\rm c)}
Department of Physics, Tokyo Metropolitan University
\break
Minami-ohsawa, Hachiohji, Tokyo 192, Japan }

\abstract{
We study nonperturbative features of QCD
using the dual Ginzburg-Landau (DGL) theory, where
the color confinement is realized through the dual Higgs mechanism
brought by QCD-monopole condensation.
The linear confinement potential appears in the QCD-monopole
condensed vacuum.
We study the infrared screening effect to the confinement potential
by the light-quark pair creation,
and derive a compact formula for the screened quark potential.
We study the dynamical chiral-symmetry breaking (D$\chi $SB)
in the DGL theory by solving the Schwinger-Dyson equation.
QCD-monopole condensation plays an essential role to D$\chi $SB.
The QCD phase transition at finite temperature
is studied using the effective potential formalism in the DGL theory.
We find the reduction of QCD-monopole condensation
and the string tension at high temperatures.
The surface tension is calculated
using the effective potential at the critical temperature.
The DGL theory predicts a large mass reduction of
glueballs near the critical temperature.
We apply the DGL theory to the quark-gluon-plasma (QGP)
physics in the ultrarelativistic heavy-ion collisions.
We propose a new scenario of the QGP formation via the
annihilation of color-electric flux tubes
based on the attractive force between them.
}

\endpage
%%%%%%%%%%%%%%%%%%% introduction %%%%%%%%%%%%%%%%%%%%%%%%%%%

\chapter{Dual Higgs Mechanism for Color Confinement}

The quantum chromodynamics (QCD) has been believed as the fundamental
theory of strong interactions
\REF\itzkson{
C.~Itzykson and J.~B.~Zuber, ``Quantum Field Theory'',
(McGraw-Hill, New York, 1985) 1.
}
\REF\cheng{
T.~P.~Cheng and L.~F.~Li,
``Gauge theory of elementary particle physics",
(Clarendon press, Oxford, 1984) 1.
}
\REF\kerson{
K.~Huang, ``Quarks, Leptons and Gauge Fields",
World Scientific, \nextline Singapore (1982) 1.
}
\REF\rothe{
M.~Creutz,``Quarks, gluons and lattices'',
(Cambridge press, 1983) 1. \nextline
H.~J.~Rothe, ``Lattice Gauge Theories", (World Scientific,
1992) 1.
}
[\itzkson-\rothe].
Because of the asymptotic freedom of QCD,
the ordinary perturbative technique is
useful for the analysis of the high-energy hadron reactions
[\itzkson-\rothe].
However, QCD in the infrared region
exhibits the nonperturbative features like
the color confinement and the dynamical
chiral-symmetry breaking (D$\chi $SB).
In particular, the color confinement is extremely unique
in the nonperturbative QCD, so that
it is difficult to find similar phenomena in the other
field of physics.
The color confinement is characterized by the
vanishing of the color dielectric constant, and squeezing of the
color electric flux [\rothe].
In this flux-tube picture, the string tension about 1GeV/fm is
one of the most important quantities on the color confinement
[\rothe].
On the other hand, D$\chi $SB belongs a wide category of the
spontaneous symmetry breaking,
and is characterized by the chiral condensate
$\langle \bar qq \rangle$, the pion decay constant $f_\pi $ and the
constituent quark mass.
For the study of the D$\chi $SB, the Schwinger-Dyson (SD) approach
\REF\higashijima{
K.~Higashijima, Phys.~Rev.~{\bf D29} (1984) 1228;
Prog.~Theor.~Phys. Suppl. {\bf 104} (1991) 1.
}
\REF\miransky{
V.~A.~Miransky, Sov.~J~.~Nucl.~Phys.~{\bf 38}(2) (1983) 280.
\nextline
V.~A.~Miransky, ``Dynamical Symmetry Breaking in Quantum Field
Theories (World Scientific, 1993) 1.
}
[\higashijima,\miransky]
provides a useful and powerful tool in terms of the
dynamical mass generation of light quarks.
The QCD phase transition is also an interesting
subject in the recent hadron physics.
\REF\kapusta{
J.~I.~Kapusta, ``Finite-Temperature Field Theory",
Cambridge University Press, Cambridge, 1988) 1.
}[\rothe,\kapusta]
It is expected that
the quark-gluon-plasma (QGP) phase appears
in the high-temperature system, which would be realized
in the ultrarelativistic heavy-ion collisions or
in the early universe.
The QGP phase is characterized as the deconfinement
and the chiral-symmetry restoration.

\section{Analogy between Superconductors and QCD vacuum}

About 20 years ago, Y.~Nambu
\REF\nambuA{
Y.~Nambu, Phys.~Rev.~{\bf D10} (1974) 4262.
}
\REF\thooftA{
G.~'t~Hooft, ``High Energy Physics", ed. A.~Zichichi
(Editorice Compositori, Bologna, 1975).
}
\REF\mandelstam{
S.~Mandelstam, Phys.~Rep.~{\bf C23} (1976) 245.
}
[\nambuA-\mandelstam]
proposed an interesting picture for the color confinement
based on the analogy between the superconductor and the QCD vacuum.
In the superconductor, magnetic field is excluded due to the
Meissner effect, which is caused by Cooper-pair condensation.
As the result, the magnetic flux is squeezed like the Abrikosov
vortex in the type II superconductor
\REF\lifshitz{
E.~M.~Lifshitz and L.~P.~Pitaevsii,
Vol.9 of Course of Theoretical Physics,
``Statistical Physics, Part 2",
(Pergamon press, Oxford, 1981) 1.
}
[\lifshitz].
On the other hand, the color-electric flux is excluded in the
QCD vacuum, and therefore the squeezed color-flux tube is formed
between color sources.
Thus, these two systems are quite similar,
and can be regarded as the dual version each other
[\kerson].
This idea is based on the duality of the gauge theories,
which was pointed out by P.A.M.~Dirac more than 50 years ago.

In this analogy, the color confinement is brought by the dual
Meissner effect originated from color-magnetic monopole condensation,
which corresponds to Cooper-pair condensation in the superconductivity.
As for the appearance of color-magnetic monopoles in QCD, 't~Hooft
\REF\thooftB{
G.~'t~Hooft, Nucl.~Phys.~{\bf B190} (1981) 455.
}
[\thooftB]
proposed an interesting idea of the abelian gauge
fixing, which is defined by the diagonalization of a
suitable gauge-dependent variable.
In this gauge, QCD is reduced into
an abelian gauge theory with magnetic monopoles, which will be
called as QCD-monopoles in order to distinguish from GUT-monopoles.
The QCD-monopoles appear from the hedgehog-like configuration
corresponding to
the nontrivial homotopy class on the nonabelian manifold,
$\pi _2({\rm SU}(N_c)/{\rm U}(1)^{N_c-1})=Z_\infty ^{N_c-1}$.
Then, the abelian gauge fixing is expected to provide the basis of
the analogy between the superconductor and the QCD vacuum.

We compare the dual Higgs mechanism in the QCD vacuum
with the ordinary Higgs mechanism in the superconductor
as shown in Fig.1.
%
%\figmark{\DHM}
%
In the superconductor, there are two kinds of degrees of freedom,
the gauge field (photon) and the matter field corresponding to
the electron and the metallic lattice, whose interaction
provides the Higgs mechanism through Cooper-pair condensation.
On the other hand, there is only the gauge field in the pure gauge QCD,
and therefore it seems difficult to find the analogous point between
these two systems.
However, in the abelian gauge, the diagonal part and the off-diagonal
part of gluons play different roles.
While the diagonal gluon behaves as the gauge field, the
off-diagonal gluon behaves as the charged matter and provides
QCD-monopoles.
Condensation of QCD-monopoles leads to mass generation
of the dual gauge field through the dual Higgs mechanism
\REF\suganumaA{
H.~Suganuma, S.~Sasaki and H.~Toki,
Nucl.~Phys.~{\bf B435} (1995) 207.
}
\REF\suzuki{
T.~Suzuki, Prog.~Theor.~Phys. {\bf 80} (1988) 929 ;
{\bf 81} (1989) 752. \nextline
S.~Maedan and T.~Suzuki, Prog.~Theor.~Phys. {\bf 81} (1989) 229.
}
[\suganumaA,\suzuki],
which is the dual version of the Higgs mechanism.
Thus, QCD can be regarded as the
dual superconductor in the abelian gauge.

In this framework, the nonperturbative QCD is mainly
described by the abelian gauge theory with QCD-monopoles,
which is called as the abelian dominance.
As for the validity of this scheme,
many recent studies based on the lattice gauge theory
have supported QCD-monopole condensation
and the abelian dominance in the maximal abelian gauge
\REF\kronfeld{
A.~S.~Kronfeld, M.~L.~Laursen, G.~Schierholz and U.~-J.~Wiese,
\nextline
Phys.~Lett.~{\bf B198} (1987) 516. \nextline
A.~S.~Kronfeld, G.~Schierholz and U.~-J.~Wiese,
Nucl.~Phys.~{\bf B293} (1987) 461.
}
\REF\yotsuyanagi{
T.~Suzuki and I.~Yotsuyanagi,
Phys.~Rev.~{\bf D42} (1990) 4257.
}
\REF\hioki{
S.~Hioki, S.~Kitahara, S.~Kiura, Y.~Matsubara,
O.~Miyamura, S.~Ohno and T.~Suzuki,
Phys.~Lett.~{\bf B272} (1991) 326.
}
\REF\shiba{
H.~Shiba and T.~Suzuki,
Nucl.~Phys.~{\bf B} (Proc.~Suppl.) {\bf 34} (1994) 182.
}
[\kronfeld-\shiba].

It is worth mentioning that the dual Higgs scheme
predicts the existence of the dual gauge field
$\vec B^\mu  \equiv(B^\mu _3,B^\mu _8)$
and the QCD-monopole $\chi _\alpha $ ($\alpha $=1,2,3)
as the relevant degrees of freedom related to the
color confinement [\nambuA,\suganumaA,\suzuki].
It can be proved that
both $\vec B^\mu $ and $\chi _\alpha $ are color-singlet, so that
they can be observed as physical states.
Since $\vec B^\mu $ and $\chi _\alpha $ appear in the gluon sector
of QCD, they are identified as glueballs, which are flavor-singlet.
The dual gauge field $\vec B^\mu $ appears as a massive
axial-vector particle.
The QCD-monopole $\chi _\alpha $ appears as a massive scalar particle.
These particles correspond to the weak vector boson
and the Higgs scalar in the electro-weak unified theory.

In this paper, we study nonperturbative features of QCD
using the dual Ginzburg-Landau theory
\REF\maedan{
S.~Maedan, Y.~Matsubara and T.~Suzuki,
Prog.~Theor.~Phys.~{\rm 84} (1990) 130. \nextline
S.~Kamizawa, Y.~Matsubara, H.~Shiba and T.~Suzuki,
Nucl.~Phys.~{\rm B389} (1993) 563.
}
[\suganumaA,\suzuki,\maedan],
which is an infrared effective theory of QCD based on
the dual Higgs mechanism by QCD-monopole condensation.
We study the color confinement, the infrared screening effect
due to the $q$-$\bar q$ pair creation and the
dynamical chiral-symmetry breaking
\REF\suganumaB{
H.~Suganuma, S.~Sasaki and H.~Toki,
Proc. of Int. Conf. on
``Quark Confinement and Hadron Spectrum",
Como, Italy, (World Scientific, 1994).
\nextline
S.~Sasaki, H.~Suganuma and H.~Toki, {\it ibid}.
\nextline
H.~Toki, H.~Suganuma and S.~Sasaki,
Nucl.~Phys.~{\bf A577} (1994) 353c.
}
\REF\sasaki{
S.~Sasaki, H.~Suganuma and H.~Toki,
preprint, RIKEN-AF-NP-172 (1994).
}
[\suganumaA,\suganumaB,\sasaki].
We also study the QCD phase transition at finite temperature
using the effective potential formalism in the DGL theory
\REF\ichie{
H.~Ichie, H.~Suganuma and H.~Toki,
preprint, RIKEN-AF-NP-176 (1994).
}
[\ichie].
Finally, we apply the DGL theory to the quark-gluon-plasma
physics in the ultrarelativistic heavy-ion collisions.

\chapter{
Dual Ginzburg-Landau Theory and Quark Potential}

\section{QCD-Monopole Condensation in DGL Theory}

The dual Ginzburg-Landau theory (DGL) theory
[\suganumaA,\suzuki]
is considered as an infrared effective theory of QCD
in the abelian gauge, and is described by
the diagonal gluon $\vec A^\mu  \equiv (A^\mu _3,A^\mu _8)$,
the dual gauge field $\vec B^\mu  \equiv (B^\mu _3,B^\mu _8)$
and the QCD-monopole field $\chi _\alpha  (\alpha =1,2,3)$
[\suganumaA,\suzuki],
$$
\eqalign{
{\cal L}_{\rm DGL}=
&-{1 \over 2n^2}
[n\cdot (\partial \wedge \vec A)]^\nu
[n\cdot ^*(\partial \wedge \vec B)]_\nu
+{1 \over 2n^2}
[n\cdot (\partial \wedge \vec B)]^\nu
[n\cdot ^*(\partial \wedge \vec A)]_\nu \cr
&-{1 \over 2n^2}
[n\cdot (\partial \wedge \vec A)]^2
-{1 \over 2n^2}
[n\cdot (\partial \wedge \vec B)]^2
+\bar q (i\partial \kern -2mm / - e \vec H \cdot \vec A \kern -2mm / -m)q \cr
&+\sum_{\alpha =1}^3[|(i\partial_\mu -g \vec \epsilon _\alpha
\cdot \vec B_\mu )\chi _\alpha |^2
-\lambda (|\chi _\alpha |^2-v^2)^2]
}
\eqn\DGLlag
$$
in the Zwanziger form
\REF\zwanziger{
D.~Zwanziger, Phys.~Rev.~{\bf D3} (1971) 880.
}
[\zwanziger], where the duality
of the gauge theory becomes manifest.
Here, $e$ is the gauge coupling constant,
$g$ is the unit magnetic charge obeying the Dirac condition
$eg=4\pi $, and $\vec \epsilon_\alpha $ denotes the relative
magnetic charge of the QCD-monopole field $\chi _\alpha $
[\suganumaA,\suzuki].
It should be noted that the magnetic charge $g \vec \epsilon _\alpha $ is
pseudoscalar because of the extended Maxwell equation,
$\nabla  \cdot {\bf H}=\rho _m$.
Hence, the dual gauge field $\vec B_\mu $ is axial-vector.
In the absence of matter fields,
one finds an exact dual relation between $\vec A_\mu $ and $\vec B_\mu $
in the field equation,
$\partial \wedge \vec B = ^* (\partial \wedge \vec A)$.

In the DGL theory, the self-interaction of the QCD-monopole field
$\chi _\alpha $ is introduced to realize QCD-monopole condensation.
When QCD-monopoles are condensed,
the dual Higgs mechanism occurs, and
the dual gauge field $\vec B_\mu $ becomes massive, $m_B= \sqrt 3 gv$.
The color-electric field is then excluded in the QCD vacuum
through the dual Meissner effect, and is
squeezed between color sources to form the hadron flux tube.
The QCD-monopole also becomes massive as $m_\chi  = 2\sqrt\lambda v$.

As for the symmetry of the DGL theory,
there is the dual gauge symmetry [U(1)$_3$$\times $U(1)$_8$]$_m$
corresponding to the local phase invariance of the QCD-monopole
field $\chi _\alpha $ [\suganumaA,\suzuki]
as well as the residual gauge symmetry [U(1)$_3$$\times $U(1)$_8$]$_e$
embedded in SU(3)$_c$.
The dual gauge symmetry leads to
the conservation of the color-magnetic flux.
In the QCD-monopole condensed vacuum,
the dual gauge symmetry [U(1)$_3$$\times $U(1)$_8$]$_m$ is
spontaneously broken due to
mass generation of the dual gauge field $\vec B_\mu $
through the dual Higgs mechanism.
Hence, the color-magnetic flux is not conserved in
the QCD-monopole condensed vacuum.
On the other hand, the residual gauge symmetry
[U(1)$_3$$\times $U(1)$_8$]$_e$
is never broken in this process [\suganumaA].

In this framework, the Dirac condition $eg=4\pi $ for the
dual gauge coupling constant $g$ is naturally
derived [\thooftB,\suganumaA]
in the same way as in the Grand Unified Theory [\cheng].
The DGL theory in the pure gauge is renormalizable,
and is not asymptotically free on $g$ in view of the
renormalization group. [See Eq.(5.2).]
Hence, asymptotic freedom is expected for the gauge coupling
constant $e$ owing to the Dirac condition.
Thus, the DGL theory qualitatively
shows asymptotic freedom on $e$ [\suganumaA,\suganumaB],
which seems a desirable feature for an effective theory of QCD.

\section{Quark Confinement Potential}

We investigate the inter-quark potential
in the quenched level using the DGL theory
[\suganumaA].
By integrating over $A_\mu$ and $ B_\mu$ in the partition
functional of the DGL theory, the current-current correlation
[\suganumaA,\suzuki] is obtained as
$$
{\cal L}_j=-{1 \over 2}\vec j_\mu D^{\mu \nu }\vec j_\nu
\eqn\JJ
$$
with the nonperturbative gluon propagator,
$$
D_{\mu \nu }={1 \over \partial^2}\{g_{\mu \nu }+(\alpha _e-1)
{\partial_\mu \partial_\nu \over \partial^2} \}
-{1 \over \partial^2}
{m_B^2 \over \partial^2+m_B^2}
{1 \over (n\cdot \partial)^2}
\epsilon ^\lambda  \ _{\mu \alpha \beta }\epsilon _{\lambda \nu \gamma \delta
}n^\alpha n^\gamma \partial^\beta \partial^\delta
\eqn\GLa
$$
in the Lorentz gauge.
Putting a static quark with color charge [\kerson,\suganumaA]
$\vec Q$ at ${\bf x}={\bf r}$
and a static antiquark with color charge
$-\vec Q$ at ${\bf x}={\bf 0}$, the quark current is written as
$
\vec j_\mu (x)=\vec Qg_{\mu 0}\{\delta ^3({\bf x}-{\bf r})
-\delta ^3({\bf x}-{\bf 0})\}.
$
We take ${\bf n} // {\bf r}$,
which is also used in the similar context of the dual string theory
[\nambuA],
because of the axial symmetry of the system
and the energy minimum condition [\suganumaA].
Otherwise, the energy of the system diverges.

We finally obtain the inter-quark potential
including the Yukawa and the linear parts [\suganumaA],
$$
V(r)=-{\vec Q^2 \over 4\pi }\cdot {e^{-m_Br} \over r}+kr,
\quad
k={\vec Q^2 m_B^2 \over 8\pi }\ln({m_B^2+m_\chi ^2 \over m_B^2})
\eqn\POT
$$
with $\vec Q^2=e^2/3$ for quarks.
Here, $m_B$ is the mass of the dual gauge field $\vec B_\mu $,
whose inverse corresponds to the cylindrical radius of the flux tube.
It should be noted that
the expression of the string tension $k$ is quite similar
to the energy per unit length of the Abrikosov
vortex in the type-II superconductor [\lifshitz].

We compare the static potential with the
phenomenological one, for example, the Cornell potential
\REF\lucha{
W.~Lucha, F.~F.~Sch\"oberl and D.~Gromes,
Phys.~Rep.~{\bf 200}, No.4
(1991) 127.
}[\lucha].
We get a good agreement as shown in Fig.2
with the choice of $e=5.5$, $m_B=0.5{\rm GeV}$
and $m_\chi =1.26{\rm GeV}$ corresponding to
$\lambda=25$ and $v=126{\rm MeV}$,
which provide $k$=1.0GeV/fm for the
string tension and the radius of the hadron flux as
$m^{-1}_B=0.4{\rm fm}$.
%
%\figmark{\QUARKPOT}
%
Thus, the linear potential responsible for the quark confinement
is reproduced in the DGL theory.

\chapter{Quark Pair Creation and Infrared Screening Effect}

In this chapter, we study the dynamical effect of light quarks
and the infrared screening effect to the confinement potential.
The dynamical effect of light quarks is important for the color
confinement, because the linear quark potential is screened
due to the $q$-$\bar q$ pair creation in the long distance
[\suganumaA].
In another words, a long hadron string can be cut through the
light $q$-$\bar q$ pair creation.
Hence, the static quark potential seems to be saturated
in the infrared region.
Such a tendency is observed in the lattice QCD with dynamical quarks
\REF\wien{
W.~Sakuler, W.~B\"urger, M.~Faber, H.~Markum, M.~M\"uller,
P.~De Forcrand, A.~Nakamura and I.~O.~Stamatescu,
Phys.~Lett.~{\bf B276} (1992) 155. \nextline
W.~B\"urger, M.~Faber, H.~Markum and M.~M\"uller,
Phys.~Rev.~{D47} (1993) 3034.
}[\wien].

\section{Quark-Pair Creation Rate in Hadron Flux Tubes}

We estimate the $q$-$\bar q$ pair creation rate
in the color-electric field inside the hadron flux tube,
which is formed between valence quarks.
The Schwinger formula
\REF\SGTA{
H.~Suganuma and T.~Tatsumi, Phys.~Lett. {\bf B269} (1991) 371.
\nextline
H.~Suganuma and T.~Tatsumi, Prog.~Theor.~Phys. {\bf 90} (1993) 379.
}
\REF\SGTB{
H.~Suganuma and T.~Tatsumi, Ann.~Phys.~(N.Y.)~{\bf 208} (1991) 470.
}
[\itzkson,\suganumaA,\SGTA,\SGTB]
for the $q$-$\bar q$ pair creation rate $w$ is given by
$$
w= -{N_f \over 2\pi ^2}
   \int_0^\infty  dp_{_T} p_{_T}
   {\rm tr}_c(eE)\ln\{1-e^{-\pi (p_{_T}^2+M^2)(eE)^{-1}}\},
\eqn\PAIRa
$$
where $M$ is the effective quark mass, and
$E$ is the external color-electric field assumed as
$E=(E_3 T_3 +E_8 T_8 )$.
It should be noted that the $q$-$\bar q$ pair creation
by the Schwinger mechanism
is a nonperturbative effect in terms of the
gauge coupling constant $e$ [\suganumaA,\SGTA].

For the $q$-$\bar q$ pair creation in the hadron flux tube,
one finds
$$
w=-{N_f \over 2\pi ^2}
   \int_0^\infty  dp_{_T} p_{_T}
   [2k \ln\{1-e^{-\pi (p_{_T}^2+M^2)/(2k)}\}
   +2 \cdot k \ln\{1-e^{-\pi (p_{_T}^2+M^2)/k}\}].
\eqn\PAIRb
$$
The first term in the bracket corresponds
to the creation of the $q$-$\bar q$ pair with
the same color as the valence quark,
and such a process contributes to the cut of the hadron flux tube
[\suganumaA].
In this case,
the effective color charge of the created $q$-$\bar q$ pair
becomes a half value due to the screening effect or the final-state
interaction
\REF\glendenning{
N.~K.~Glendenning and T.~Matsui,
Phys.~Rev.~{\bf D28} (1983) 2890; \nextline
Phys.~Lett.~{\bf B141} (1984) 419.
}
[\suganumaA,\glendenning],
so that $2k$ in the first term of Eq.{\PAIRb} is reduced to $k$.
On the other hand, the second term in Eq.{\PAIRb} describes the
creation of the $q$-$\bar q$ pair with different color
from the valence quarks, so that this contribution is
less important to the screening of the quark potential [\suganumaA].

Thus, the $q$-$\bar q$ pair creation rate relevant
to the screening effect is given by
$$
w_{\rm sc}=-{N_f \over 2\pi ^2}k
   \int_0^\infty  dp_{_T} p_{_T}
     \ln\{1-e^{-\pi (p_{_T}^2+M^2)/k}\}
 \equiv
   \int_0^\infty  dp_{_T} w(p_{_T}),
\eqn\PAIRc
$$
where $w(p_{_{T}})$ is the creation rate of the $q$-$\bar q$ pair with
the transverse momentum $p_{_T}$.
The expectation value of the energy of the created
$q$-$\bar q$ pair is estimated by
$
\langle 2 E_q \rangle \simeq {1 \over w_{\rm sc}}
          \int_0^\infty  dp_{_T} w(p_{_T}) \cdot 2
          (p_{_T}+M^2)^{1/2}
          \simeq 850{\rm MeV}
%\eqn\EVEV
$
[\suganumaA]
for $k = 1.0 {\rm GeV/fm}$ and $M = 350{\rm MeV}$.
Since the energy $\langle 2 E_q \rangle $ is supplied
by the missing length of the hadronic string,
the infrared screening length $R_{\rm sc}$ satisfies
$k R_{\rm sc} \simeq \langle 2 E_q \rangle $.
Hence, one obtains $R_{\rm sc} \simeq 1{\rm fm}$ [\suganumaA],
which corresponds to a typical value of the hadron size.

\section{Infrared Screening Effect to Confinement Potential}

The hadronic string becomes unstable against
the $q$-$\bar q$ pair creation
when the distance between the valence
quarks becomes larger than $R_{\rm sc}$.
This means the vanishing of the strong correlation between the
valence quarks in the infrared region, so that
the corresponding infrared cutoff,
$a \simeq R_{\rm sc}^{-1} \simeq 200{\rm MeV}$,
should be introduced to the system [\suganumaA].
Taking account of such an infrared screening effect,
we introduce the corresponding infrared cutoff
$a$ to the nonperturbative gluon propagator {\GLa} by replacing
${1 \over (n \cdot k)^2} \rightarrow {1 \over (n \cdot k)^2 +a^2}$
[\suganumaA],
$$
D_{\mu \nu}^{sc}=-{1 \over k^2}
\left\{ g_{\mu \nu }+(\alpha_e-1)
{k_\mu k_\nu \over k^2} \right\}
-{1 \over k^2}
{m_B^2 \over k^2-m_B^2} \cdot
{\epsilon ^\lambda  \ _{\mu \alpha \beta }\epsilon _{\lambda \nu \gamma \delta
}n^\alpha n^\gamma k^\beta k^\delta
\over (n \cdot k)^2+a^2},
\eqn\PROP
$$
because the non-local factor ${1 \over (n \cdot k)^2}$
provides the strong and long-range correlation as
the origin of the confinement potential [\suganumaA].
Here, this gluon propagator keeps the residual gauge symmetry.
Such a disappearance of the infrared double pole in
the gluon propagator in the DGL theory can be qualitatively
shown by considering the polarization diagram of quarks.

Using the gluon propagator {\PROP},
we obtain a compact formula [\suganumaA] for the quark
potential including the infrared screening effect due to the
$q$-$\bar q$ pair creation,
$$
V_{\rm sc}(r)
=-{\vec Q^2 \over 4\pi }\cdot {e^{-m_Br}\over r}
+k \cdot {1-e^{-ar} \over a},
\eqn\POTc
$$
which exhibits the saturation
for the longer distance than $a^{-1} \simeq 1{\rm fm}$.
This formula for the screened quark potential
is used for the phenomenological analysis of the hadron decay
\REF\chao{
K.-T.~Chao, Y.-B.~Ding and D.-H.~Qin, Commun.~Theor.~Phys.~{\bf 18}
(1992) 321.
}
[\chao],
and seems consistent with the recent studies of the
lattice QCD including light dynamical quarks
[\wien].

\chapter{Dynamical Chiral-Symmetry Breaking}

We now study the dynamical chiral-symmetry
breaking (D$\chi $SB) in the dual Ginzburg-Landau (DGL) theory
[\suganumaA,\sasaki],
considering the relation between the color confinement
and D$\chi $SB, which is suggested by the lattice QCD with dynamical
quarks [\rothe].
We investigate D$\chi $SB in terms of the dynamical quark-mass generation
in the QCD-monopole condensed vacuum
by using the Schwinger-Dyson (SD) equation for massless quarks
[\suganumaA,\sasaki],
$$
S^{-1}_q(p)={p \kern -2mm /}+ \int {d^4k \over i(2\pi )^4}
\vec Q^2 \gamma ^\mu S_q(k)\gamma ^\nu
D_{\mu \nu }^{\rm sc}(k-p).
\eqn\SDEa
$$
Here, we use the gluon propagator $D_{\mu \nu }^{\rm sc}$
including the nonperturbative effects on the
color confinement and the infrared screening
in the presence of light quarks [\suganumaA].
The quark propagator $S_q(p)$ is assumed as
$
S_q(p)^{-1}={p \kern -2mm /} -M(-p^2)+i\epsilon .
$

Taking the trace and making the Wick rotation in the $k_0$-plane,
we obtain the SD equation in the Euclidean metric,
$$
M(p^2)= \int{d^4k \over (2\pi )^4} \vec Q^2 {M(k^2) \over k^2+M^2(k^2)}
D_\mu ^{\mu {\rm sc}}(k-p),
\eqn\SDEb
$$
where the trace of the gluon propagator {\PROP} is given as
$$
D_\mu ^{\mu {\rm sc}}(k) =
 {1 \over (n \cdot k)^2+a^2} \cdot {1 \over k^2} \cdot
 {2m_B^2 \over k^2+m_B^2}\{k^2-(n \cdot k)^2 \}
 +{3+\alpha _e \over k^2}
\eqn\GP
$$
in the Lorentz gauge.
After performing the angular integration,
we obtain the final expression for the SD equation [\suganumaA],
$$
\eqalign{
M(p^2)
 &=\int_0^\infty {dk^2 \over 16\pi ^2} {\vec Q^2 M(k^2)
 \over k^2+M^2(k^2)} \biggl(
 {4k^2  \over   k^2+p^2+m_B^2 + \sqrt{(k^2+p^2+m_B^2)^2-4k^2p^2} }
 \cr
 &+{(1+\alpha _e)k^2 \over {\rm max}(k^2, p^2)}
 +{1 \over \pi p_{_T}}
 \int_{-k}^k dk_n {1 \over \tilde k_n^2+a^2} \cr
 &\times [
 (m_B^2-a^2) \ln \{ { {\tilde k_n^2+(k_{_T}+p_{_T})^2+m_B^2
 \over \tilde k_n^2+(k_{_T}-p_{_T})^2+m_B^2} } \}
 +a^2   \ln \{ { {\tilde k_n^2+(k_{_T}+p_{_T})^2
 \over \tilde k_n^2+(k_{_T}-p_{_T})^2} } \}     ] \biggr)
}
\eqn\SDEc
$$
with $\tilde k_n \equiv k_n-p_n$ and $k_{_T} \equiv (k^2-k_n^2)^{1/2}$.
Here, one finds $\vec Q^2=e^2/3$ for quarks.
Since, the integrand in the SD equation
is non-negative, the existence of the nontrivial solution
can be expected.

In solving the SD equation, we use the Higashijima-Miransky
approximation
[\higashijima,\miransky]
with a hybrid type of the running coupling constant,
$$
\tilde e=e({\rm max}\{p^2, k^2\}), \quad
e^2(p^2)=
{48\pi ^2 (N_c+1) \over (11N_c-2N_f)
\ln\{(p^2+p^2_c)/\Lambda ^2_{\rm QCD}\}}.
\eqn\HMA
$$
Here, $p_c$ is defined as
$p_c\equiv \Lambda _{\rm QCD}\exp[{24 \pi ^2 \over e^2}\cdot
{N_c+1 \over 11N_c-2N_f}]$ with $e = e(0)$ [\suganumaA].
This ansatz naturally connects to the asymptotic freedom of the
running coupling at large momentum.
The coupling constant at low energy, $e(p^2\sim 0) \simeq e$,
controls the strength of the linear confinement potential.

We show in Fig.3 the quark mass function $M(p^2)$ with
$e$=5.5 and $a =80{\rm MeV}$.
%
%\figmark{\QUARKMASS}
%
The QCD scale parameter is set to a realistic value
$\Lambda _{\rm QCD}=200{\rm MeV}$ in the $\bar{\rm MS}$ scheme.
In order to see the effect of QCD-monopole condensation,
we vary the mass of the dual gauge field, $ m_B $.
There is no non-trivial solution for the case with small
$m_B < 300{\rm MeV}$.
A non-trivial solution is barely obtained at $ m_B=300{\rm MeV}$,
and $M(p^2)$ increases rapidly with $ m_B $ as shown in Fig.3.
Thus, QCD-monopole condensation provides a crucial
contribution to D$\chi $SB [\suganumaA].

We further examine the result for $M(p^2)$ as shown in Fig.4
for $ m_B=0.5{\rm GeV}$, which was used
for the argument of the confinement potential in chapter 2.
%
%\figmark{\QUARKPROP}
%
The quark mass function $M(p^2)$ in the space-like region
is directly obtained from the SD equation.
We extrapolate $M(p^2)$ into the time-like region
using a polynomial function as a simulation
of the analytic continuation.
This curve does not satisfy
the on-shell condition $M^2(p^2)+p^2=0$ ($p_\mu $: Euclidean momentum),
and hence the quark propagator does not have a physical pole
[\suganumaA,\sasaki].
This may indicate the light-quark confinement.

We find that monopole condensation makes the slope of $M(p^2)$
around $p^2 \simeq 0$ larger as can be seen
by comparing the $m_B=0$ case with the $m_B \ne 0$ case.
This tendency can be physically explained as follows.
The strong confining force between $q$ and $\bar q$ appears
due to QCD-monopole condensation, and this attractive force
should promote $q$-$\bar q$ pair condensation
[\suganumaA,\sasaki].
Since such an effect of the confining force becomes stronger
in the infrared region $p^2 \simeq 0$,
the corresponding dynamical mass generation of quarks
is much enhanced there,
which provides the large slope of $M(p^2)$ at $p^2 \simeq 0$.
Thus, the light-quark confinement or
the absence of physical poles of quarks
is achieved due to this modification of the quark propagator
in the infrared region by the confinement effect
[\suganumaA,\suganumaB,\sasaki].

We also calculate the other quantities related to D$\chi $SB
from the solution of the SD equation.
The constituent quark mass in the infrared region
is found to be $M(0)$=348MeV.
The quark condensate is obtained as
$\langle \bar {q}q \rangle =-(229{\rm MeV})^3$.
The pion decay constant is also calculated as
$f_\pi $=83.6MeV using the Pagels-Stoker formula
\REF\ps{
H.~Pagels and S.~Stoker, Phys.~Rev.~{\bf D20} (1979) 2947;
{\bf D22} (1980) 2876.
}[\ps].
These values are to be compared with the standard values;
$M$(0)=350 MeV, $\langle \bar{q}q \rangle
=-(225 \pm{\rm50} {\rm MeV})^3$ and $f_{\pi}=93{\rm MeV}$.
Thus, several relevant quantities on D$\chi $SB
can be reproduced in the DGL theory [\suganumaA,\sasaki].

\chapter{QCD Phase Transition at Finite Temperature}

\section{Effective Potential Formalism}

In this chapter,
we study the change of the QCD vacuum at finite temperature
using the dual Ginzburg-Landau (DGL) theory
in terms of QCD-monopole condensation
\REF\monden{ H.~Monden T.~Suzuki and Y.~Matsubara,
Phys.~lett.~{\bf B 294} (1992) 100.
}
[\ichie,\monden].
To concentrate on the confinement properties, we study at the
quenched level [\rothe],
where the quark degrees of freedom are frozen.
In this case, we can drop the quark term in the DGL Lagrangian
and perform integration over the gauge field $A_\mu $.
Hence, we obtain the partition functional as [\ichie]
$$
  Z[J] = \int {\cal D}{\chi_{\alpha}}{\cal D}{\vec{B}_{\mu}}
\exp{\left( i\int d^4x \{{\cal{L}}_{\rm DGL}
-J\sum_{\alpha=1}^3|\chi_\alpha|^2\} \right) },
\eqn\Zj
$$
where ${\cal L}_{\rm DGL}$ has a simple form,
$$
 {\cal L}_{\rm DGL}= - {1 \over 4}
(\partial_{\mu}\vec{B}_{\nu}-\partial_{\nu}\vec{B}_{\mu})^2 +
 \sum_{\alpha=1}^3[|(i\partial_{\mu}-g\vec{\epsilon_{\alpha}}
\cdot\vec{B}_{\mu})
\chi_{\alpha}|^2 - \lambda(|\chi_{\alpha}|^2-v^2)^2].
\eqn\La
$$

Here, we have introduced the quadratic source term
\REF\ring{P.~Ring and P.~Schuck, ``The Nuclear Many-Body Problem'',
(Springer-Verlag, New York, 1980) 1.
}
[\ichie,\ring]
instead of the linear source term, which is commonly used.
As is well-known in the $\phi^4$ theory [\cheng,\kapusta],
the use of the linear source term
leads to an imaginary mass of the scalar field $\chi _\alpha $
in the negative-curvature region of the
classical potential, and therefore
the effective action cannot be obtained there due to
the appearance of ``tachyons".
In this respect, there is an extremely advanced point in
the use of the quadratic source term [\ring],
because the mass of the scalar field $\chi _\alpha $ is always real even
in the negative-curvature region of the classical potential
owing to the contribution of the source $J$
to the scalar mass. [See Eq.(5.4).]
Then, one obtains the effective action for the
whole region of the order parameter without any difficulty
of the imaginary-mass problem.
Since this method with the quadratic source term is quite general,
it is convenient to formulate the non-convex effective potential
in the $\phi ^4$ theory, the linear $\sigma $ model
or the Higgs sector in the unified theory [\itzkson-\kerson].

The effective potential at finite temperature,
which physically corresponds to the thermodynamical potential,
is then obtained as [\ichie]
$$
\eqalign{
V_{\rm eff}(\bar \chi ;T) =   3 \lambda ( \bar \chi^2 - v^2 )^2
          &+ 3 {T \over \pi^2} \int_0^\infty  dk k^2 \ln{
           \left(  1 - e^{ - \sqrt{ k^2 + m_B^2}/T }
           \right)  } \cr
          &+ {3 \over 2} {T \over \pi^2}\int_0^\infty  dk k^2 \ln{
           \left(  1 - e^{ - \sqrt{ k^2 + m_{\chi}^2}/T }
           \right)  }.
}
\eqn\Vb
$$
Here, the masses of the QCD-monopole and the dual gauge field
depend on the QCD-monopole condensate $\bar \chi $,
$$
   m_\chi ^2(\bar \chi ) = 2\lambda (3 \bar \chi ^2-v^2) + J(\bar \chi ) =
4\lambda \bar \chi ^2,
   \hbox{\quad} m_B^2(\bar \chi ) = 3 g^2 \bar \chi ^2.
\eqn\Ma
$$

We show in Fig.5  the effective potential
$V_{\rm eff}(\bar \chi ;T)$
as a function of the QCD-monopole condensate $\bar \chi $.
%
%\figmark{\EFFPOT}
%
The (local-)minimum point, $\bar \chi _{\rm phys}(T)$,
of $V_{\rm eff}(\bar \chi ;T)$
corresponds to the physical (meta-)stable vacuum state.
At $T=0$, one minimum appears at a finite $\bar \chi $,
which corresponds to the QCD-monopole condensed phase.
As the temperature increases, the minimum moves toward a small
$\bar \chi $ value, and the second minimum appears at $\bar \chi =0$
above the lower critical temperature $T_{\rm low} \simeq 0.38$GeV,
which is analytically obtained
using the high-temperature expansion [\kapusta,\ichie,\monden],
$$
T_{\rm low}=2v \sqrt{6\lambda  \over 2\lambda +3g^2}.
\eqn\LCT
$$
The potential values at the two minima become equal at
the thermodynamical critical temperature
$T_c \simeq 0.49$GeV.
The trivial vacuum ($\bar \chi =0$) becomes stable above $T_c$.
Thus, this phase transition is of the first order.

Here, we consider the possibility of the temperature dependence
on the parameters ($\lambda $,$v$) in the DGL theory.
The critical temperature, $T_c$ = 0.49 GeV, seems much larger
than the lattice QCD result, $T_c \simeq 0.2$GeV [\rothe].
Here, the self-interaction of $\chi _\alpha $
is introduced phenomenologically in the DGL Lagrangian,
and it would be reduced at high $T$ according to
the asymptotic freedom behavior of QCD.
Here, we use a simple ansatz for the $T$-dependence
on $\lambda $ [\ichie],
$
\lambda (T) \equiv \lambda  ( 1 - \alpha  T/T_c),
$
where $\alpha $ is determined as $\alpha =0.96$
so as to reproduce $T_c=0.2$GeV. (We take $\lambda (T)=0$ for $T>T_c/\alpha $.)
The qualitative behavior is the same as in the
above argument with a constant $\lambda $.
We find a first-order phase transition again.
Then, a large reduction of the
self-interaction among QCD monopoles is
expected near the critical temperature $T_c$:
$\lambda (T \simeq T_c) \simeq 1$
is considerably smaller than $\lambda (T=0)=25$.
Such a large reduction of $\lambda (T)$ near $T_c$
may be checked using the lattice QCD.

When light dynamical quarks are included,
the chiral-symmetry restoration is also expected in the DGL theory
as well as the deconfinement phase transition
at the critical temperature,
because QCD-monopole condensation is essential
for D$\chi $SB as demonstrated in the previous chapter.

\section{Glueball Mass, String Tension and Surface Tension}

We investigate the masses of the
dual gauge field $\vec B_\mu $ and the QCD-monopole field
$\tilde \chi _\alpha $ at finite temperatures.
Here, $\vec B_\mu $ and $\tilde \chi _\alpha $ would appear as
the color-singlet glueball field with $1^+$ and $0^+$, respectively
[\suzuki,\suganumaB].
In Fig.6, we show the glueball masses
$m_B(T)$ and $m_\chi (T)$ using variable $\lambda (T)$.
%
%\figmark{\GBMASS}
%
It is worth mentioning that $m_B(T)$ and $m_\chi (T)$
drop down to $m_B, m_\chi  \sim T_c$ ($\simeq$ 0.2GeV)
from $m_B, m_\chi  \sim$ 1 GeV near the critical temperature $T_c$.
In other words,
the QCD phase transition occurs at the temperature satisfying
$m_B, m_\chi  \simeq T$.
Thus, our result predicts a large reduction
of the glueball masses, $m_B$ and $m_\chi $, near the
critical temperature $T_c$.

This result would be natural because of the following argument.
In general, the thermodynamical factor
$
1 / (e^{\omega _n/T} \pm 1 )
$
for the single-particle energy $\omega _n$
becomes relevant only for $\omega _n \lsim T$.
Hence, one may guess relatively high critical temperature
$T_c \gsim 1{\rm GeV}$ in the pure gauge QCD,
because only heavy glueballs appear as the elementary
excitations there.
However, the lattice QCD shows $T_c \simeq 200{\rm MeV} \ll 1{\rm GeV}$.
This discrepancy would be solved by a large
reduction of the glueball mass near the critical temperature,
as was demonstrated in the DGL theory [\ichie].
Similar glueball-mass reduction is also
suggested by the thermodynamical studies
based on the lattice QCD data
\REF\karsch{
J.~Engels, F.~Karsch, H.~Satz and I.~Montvay,
Phys.~Lett.~{\bf B102} (1981) 332.
}[\karsch].

We investigate the string tension $k(T)$ at finite temperatures.
The string tension $k(T)$ is obtained as shown in Fig.7
by using Eq.{\POT}.
%
%\figmark{\STRIN}
%
In the case of variable $\lambda (T)$,
the string tension $k(T)$ decreases rapidly with temperature,
and $k(T)$ drops down to zero around $T_c = $ 0.2 GeV.
Hence, one expects a rapid change of
the masses and the sizes of the quarkonia
according to the large reduction of $k(T)$
at high temperatures.
We plot also the lattice QCD data in the pure gauge
\REF\gao{ M.~Gao, Nucl.~Phys.~{\bf B9} (Proc. Suppl.) (1988) 368.}
[\gao] by black dots.
One finds that the variable $\lambda (T)$ case reproduces the lattice
QCD data.

We can also estimate the surface tension $\sigma $
between the confinement and deconfinement phases
using the effective potential at $T_c$ in the DGL theory.
We show in Fig.8 the effective potential
$V_{\rm eff}(\bar \chi ;T_c)$ for the variable $\lambda (T)$.
%
%\figmark{\EPCR}
%
There are two minima at $\bar \chi  = 0, \bar \chi _c$ in
$V_{\rm eff}(\bar \chi ;T_c)$, and we choose
the origin of the energy as
$V_{\rm eff}(0;T_c)=V_{\rm eff}(\bar \chi _c;T_c)=0$ for simplicity.
The mixed phase includes both the confinement phase
($\bar \chi =\bar \chi _c$) and the deconfinement phase ($\bar \chi =0$).
Here, we set the boundary surface in the mixed phase
on $xy$-plane ($z=0$) in order to estimate the surface tension.
In this case, the system depends on $z$-coordinate only,
and the boundary condition is given as
$$
\bar \chi (z=-\infty )=0, \quad \bar \chi (z=\infty )=\bar \chi _c.
\eqn\BCz
$$
The surface tension $\sigma $ in the DGL theory
is estimated as
$$
\sigma  \simeq \int_{-\infty }^\infty  dz  \left\{
3 \left({d \bar \chi (z) \over dz} \right)^2
+V_{\rm eff}[\bar \chi (z);T_c] \right\},
\eqn\SurT
$$
where $\bar \chi (z)$ satisfies the boundary condition {\BCz}.

The figure of $V_{\rm eff}(\bar \chi ;T_c)$
($0 \le \bar \chi  \le \bar \chi _c$)
is approximated as a sine curve,
$$
V_{\rm eff}(\bar \chi ;T_c)
\simeq  {h \over 2}
\{1-\cos (2\pi  \bar \chi  / \bar \chi _c)\}
\quad
(0 \le \bar \chi  \le \bar \chi _c)
\eqn\SiN
$$
with $h$ the ``height'' of $V_{\rm eff}(\bar \chi ;T_c)$.
Then, the field equation of $\bar \chi (z)$
can be solved analytically like the sine-Gordon equation
\REF\rajaraman{
R.~Rajaraman, ``Solitons and Instantons",
(North-Holland, Amsterdam, 1982) 1.
}[\rajaraman],
$$
\bar\chi (z) \simeq {2\sqrt{6} \over 3} \tan^{-1} e^{z/\delta },
\quad
\delta  \equiv { \sqrt{3} \over \pi } \bar \chi _c/\sqrt{h},
\eqn\SurTA
$$
where $\delta $ denotes the thickness of the boundary
between the two phases.
We obtain a simple formula for the surface tension $\sigma $,
$$
\sigma  \simeq {4\sqrt{3} \over \pi } \sqrt{h} \bar \chi _c.
\eqn\SurTC
$$

One finds $\bar \chi _c \simeq 0.49{\rm fm}^{-1}$ and
$h \simeq 0.026 {\rm fm}^{-4}$ from
$V_{\rm eff}(\bar \chi ;T_c)$ in Fig.8.
Hence, the surface tension is estimated as
$\sigma  \simeq (112{\rm MeV})^3$,
and the thickness of the border between the two phases
is $\delta  \simeq 1.68{\rm fm}$.
Since the above estimation has been done in the quenched level,
the obtained results are to be compared with
the lattice QCD data in the quenched level, e.g.
$\sigma ^{1/3} \sim 60 {\rm MeV}$
\REF\iwasaki{
Y.~Iwasaki, K.~Kanaya and L.~Karkkainen,
Phys.~Rev.~{\bf D49} (1994) 3540.
}
[\iwasaki].

\chapter{Application to Quark-Gluon-Plasma Physics}

Finally, we apply the DGL theory to the
quark-gluon-plasma (QGP) physics in ultrarelativistic
heavy-ion collisions.
In a modern picture of the QGP formation
\REF\gatoff{
G.~Gatoff, A.~K.~Kerman and T.~Matsui, Phys.~Rev.~{\bf D36} (1987) 114.
}
[\SGTA,\glendenning,\gatoff],
many color-electric flux tubes are
formed between heavy ions
immediately after the collision.
In this pre-equilibrium stage,
there occurs $q$-$\bar q$ pair creation violently
inside tubes by the Schwinger mechanism [\SGTA,\glendenning].
During this process, the energy of the color-electric field
turns into that of the stochastic kinetic motion of quarks (and gluons).
The energy deposition and the thermalization thus occur.
So far, many studies has been done to the
properties of the QGP phase [\rothe,\kapusta,\ichie]
in the equilibrium stage,
however, the pre-equilibrium stage is also
important in terms of the QGP formation [\SGTA].

For the study of the QGP formation,
the DGL theory would provide a useful method,
because it describes the properties
of the color-electric flux tube, which are important
in the pre-equillibrium system just after the
ultrarelativistic heavy-ion collisions.
These would be lots of flux tubes overlapping
in the central region between heavy ions,
when the energy of the collision is enough high.
Hence the interaction between the flux tubes
is very important in this case, although dynamics of the
flux tubes is neglected in most studies of the QGP formation.

There are several kinds of flux tubes in the QCD system.
Each flux tube is characterized by the
color charge $\vec Q$ [\kerson,\suganumaA] at its one end.
To classify sorts of the flux tube,
we call the flux tube with a red quark ($R$) at its one end
as ``$R$-$\bar R$ flux tube'', and so on.
In this case, the ``direction'' of the color-electric flux
in the flux tube should be distinguished.
For instance, $\bar R$-$R$ flux tube is different
from $R$-$\bar R$ flux tube in terms of the flux direction.

We study the interaction between two color-electric flux tubes
using the DGL theory.
The color-electric charges at one end of the flux tubes are
denoted by $\vec Q_1$ and $\vec Q_2$.
We idealize the system as two sufficiently long flux tubes,
and neglect the effect of their ends.
We denote by $d$ as the distance between the two flux tubes.
For $d \gg m_\chi ^{-1}$,
the interaction energy per unit length in the two flux tube system
is estimated as
$$
E_{\rm int} \simeq {8\pi  \vec Q_1 \cdot \vec Q_2  \over e^2}
m_B^2 K_0(m_Bd),
\eqn\VVINT
$$
where $K_0(x)$ is the modified Bessel function.
Here, we have used the similar calculation
on the Abrikosov vortex in the type-II superconductor [\lifshitz].

As shown in Fig.9, there are two interesting cases
on the interaction between two color-electric flux tubes.
%
%\figmark{\FTAN}
%

\item{\rm (a)}
For the same flux tubes with opposite flux
direction (e.g. $R$-$\bar R$ and $\bar R$-$R$),
one finds $\vec Q_1=-\vec Q_2$ i.e.
$\vec Q_1 \cdot \vec Q_2=-e^2/3$, so that
these flux tubes are attracted each other.
It should be noted that they would be annihilated into
dynamical gluons in this case.

\item{\rm (b)}
For the different flux tubes satisfying
$\vec Q_1 \cdot \vec Q_2<0$
(e.g. $R$-$\bar R$ and $B$-$\bar B$),
one finds $\vec Q_1 \cdot \vec Q_2=-e^2/6$,
so that these flux tubes are attractive.
In this case, they would be unified into a single flux
tube (similar to $G$-$\bar G$ flux tube).

Based on the above calculation,
we propose a new scenario on the QGP formation via
the annihilation of the color-electric flux tubes.
When the flux tubes are sufficiently dense in the central region
just after ultrarelativistic heavy-ion collisions,
many flux tubes are annihilated or unified.
During the annihilation process of the flux tubes,
lots of dynamical gluons (and quarks) would be created.
Thus, the energy of the flux tubes turns into that of the
stochastic kinetic motion of gluons (and quarks).
The thermalization is achieved through the stochastic
gluon self-interaction, and finally the hot QGP would be created.
Here, the gluon self-interaction in QCD plays an
essential role to the thermalization process,
which is quite different from the photon system in QED.

In more realistic case, both the quark-pair creation and the flux-tube
annihilation would take place at the same time.
For instance, the flux tube breaking [\SGTA,\glendenning,\gatoff]
would occur before the flux tube annihilation for the
dilute flux tube system.
On the contrary,
in case of the extremely high energy collisions,
these would be lots of flux tubes overlapping
in the central region between heavy ions,
and therefore the flux tube annihilation
should play the dominant role in the QGP formation.
In any case, the DGL theory would provide a calculable method
for dynamics of the color-electric flux tubes in the QGP formation.

\chapter{Summary and Concluding Remarks}

We have studied nonperturbative features of QCD
using the dual Ginzburg-Landau (DGL) theory.
The dual Higgs mechanism brought by QCD-monopole condensation leads
to the color confinement in the DGL theory.
The linear confinement potential has been derived
in the QCD-monopole condensed vacuum.

We have studied the infrared screening effect to
the confinement potential by the light-quark pair creation.
By introducing the corresponding infrared cutoff,
we have derived a compact formula for the screened quark potential.

We have studied the dynamical chiral-symmetry breaking (D$\chi $SB)
in the DGL theory by solving the Schwinger-Dyson equation.
We have found that
QCD-monopole condensation plays an essential role to D$\chi $SB.
The quark propagator is largely modified by
the confinement effect in the infrared region,
which would leads to the absent of physical poles of quarks.

We have studied the QCD phase transition at finite temperature
using the effective potential formalism in the DGL theory.
Here, we have proposed the use of the quadratic source term
as a general powerful method in the $\phi ^4$-like theory.
We have found the reduction of QCD-monopole condensation
and the string tension at high temperatures.
The surface tension has been calculated
using the effective potential at the critical temperature.
In particular, the DGL theory predicts a large mass reduction of
glueballs, the QCD-monopole and the dual gauge field, near the
critical temperature.

Finally, we have applied the DGL theory to the quark-gluon-plasma (QGP)
physics in the ultrarelativistic heavy-ion collisions.
Based on the attractive force between color-electric flux tubes,
we have proposed a new scenario of the QGP formation via
the annihilation of the flux tubes.

One of the authors (H.S.) is supported by the Special
Researchers' Basic Science Program at RIKEN.

\refout

\endpage

%%%%%%%%%%%%%%%%%%% Figure Captions %%%%%%%%%%%%%%%%%%%%%%%%%%%%
\centerline{\fourteenpoint Figure Captions}

\vskip0.5cm

\item{Fig.1.}
Comparison between the superconductor and
the QCD vacuum in the abelian gauge.
The off-diagonal gluon behaves as the charged matter field
in the abelian gauge, and contributes to the appearance of
QCD-monopoles.
The dual Meissner effect is brought by
QCD-monopole condensation, which is the dual version of
the Meissner effect caused by Coopr-pair condensation
in the superconductivity.

\item{Fig.2.}
The static quark potential $V(r)$ in the
dual Ginzburg-Landau theory.
The dashed curve denotes the Cornell potential.

\item{Fig.3.}
The dynamical quark mass $M(p^2)$
as a function of the
Euclidean momentum squared $p^2$ for $m_B$=300, 400 and 500 MeV.

\item{Fig.4.}
The dynamical quark mass squared $M^2(p^2)$
as a function of $p^2$.
The dotted straight line denotes the on-shell state.

\item{Fig.5.}
The effective potentials at various temperatures
as functions of the QCD-monopole condensate $\bar \chi $.
The crosses denote their minima.

\item{Fig.6.}
The glueball masses at finite temperatures :
$m_B(T)$ and $m_\chi (T)$.
A large reduction of these masses is found near the critical
temperature.
The phase transition occurs at the temperature satisfying
$m_B, m_\chi  \simeq T$, which is denoted by the dotted line.

\item{Fig.7.}
The string tensions $k(T)$ as functions of the
temperature $T$ for a constant $\lambda$ and a variable $\lambda (T)$.
The lattice QCD results in the pure gauge
in Ref.[\gao] are shown by black dots.

\item{Fig.8.}
The effective potential $V_{\rm eff}(\bar \chi ;T_c)$
at the critical temperature $T_c$.
There are two minima at $\bar \chi =0, \bar \chi _c$
in $V_{\rm eff}(\bar \chi ;T_c)$, and
$h$ denotes the height of $V_{\rm eff}(\bar \chi ;T_c)$.
The dashed curve is an approximate sine one as Eq.(5.8).

\item{Fig.9.}
The annihilation process of the color-electric flux tubes
during the QGP formation in ultrarelativistic
heavy-ion collisions.
(a) The same flux tubes with opposite flux direction are attracted
each other, and are annihilated into dynamical gluons.
(b) The different flux tubes (e.g. $R$-$\bar R$ and $B$-$\bar B$)
are attractive, and are unified into a single flux tube.

\end